\documentclass[prb,twocolumn,showpacs,preprintnumbers,amsmath,amssymb,superscriptaddress,floatfix]{revtex4}

\usepackage{graphicx}
\usepackage{dcolumn}
\usepackage{bm}
\usepackage{here} 
\usepackage{epsf}
\usepackage{color}

\begin{document}


\title{Soft X-ray Angle Resolved Photoemission with Micro Positioning Techniques \\for Metallic V$_2$O$_3$}


\author{H. Fujiwara}
\email{fujiwara@mp.es.osaka-u.ac.jp}
\author{T. Kiss}
\author{Y. K. Wakabayashi}
\affiliation{Graduate School of Engineering Science, Osaka University, Toyonaka, Osaka 560-8531, Japan}
\author{Y. Nishitani}
\affiliation{Faculty of Science and Engineering, Konan University, Okamoto 8-9-1, Kobe, Hyogo, 658-8501, Japan}
\author{T. Mori}
\author{Y. Nakata}
\author{S. Kitayama}
\author{K. Fukushima}
\author{S. Ikeda}
\author{H. Fuchimoto}
\author{Y. Minowa}
\affiliation{Graduate School of Engineering Science, Osaka University, Toyonaka, Osaka 560-8531, Japan}
\author{S.-K. Mo}
\affiliation{Randall Laboratory of Physics, University of Michigan, Ann Arbor, Michigan 48109, USA}
\affiliation{Advanced Light Source, Lawrence Berkeley National Laboratory, Berkeley, California, 94720, USA}
\author{J. D. Denlinger}
\affiliation{Advanced Light Source, Lawrence Berkeley National Laboratory, Berkeley, California, 94720, USA}
\author{J. W. Allen}
\affiliation{Randall Laboratory of Physics, University of Michigan, Ann Arbor, Michigan 48109, USA}
\author{P. Metcalf}
\affiliation{Department of Physics, Purdue University, West Lafayette, Indiana 47907, USA}
\author{M. Imai}
\author{K. Yoshimura}
\affiliation{Department of Chemistry,Graduate School of Science, Kyoto University, Oiwake town, Kitashirakawa, Sakyo, Kyoto, 606-8502, Japan}
\author{S. Suga} 
\affiliation{Institute of Scientific and Industrial Research, Osaka University, Ibaraki, Osaka, 567- 0047, Japan}
\author{T. Muro}
\affiliation{Japan Synchrotron Radiation Research Institute (JASRI), 1-1-1 Kouto, Sayo, Hyogo 679-5198, Japan}
\author{A. Sekiyama} 
\affiliation{Graduate School of Engineering Science, Osaka University, Toyonaka, Osaka 560-8531, Japan}

\date{\today}

\begin{abstract}
We have performed soft-X-ray angle resolved photoemission for metallic V$_2$O$_3$. Combining a micro focus beam (40 x 65 ${\mu}$m$^2$) and micro positioning techniques with a long working distance microscope, we have succeeded in observing band dispersions from tiny cleavage surfaces with typical size of the several tens of ${\mu}$m. The photoemission spectra show a clear position dependence reflecting the morphology of the cleaved sample surface. By selecting high quality flat regions on the sample surface, we have succeeded in band mapping using both photon-energy and polar-angle dependences, opening the door to three-dimensional ARPES for typical three dimensional correlated materials where large cleavage planes are rarely obtained.
\end{abstract}


\maketitle
%

\section{INTRODUCTION}

Soft-X-ray angle resolved photoemission (ARPES) is a powerful tool for investigating Fermi
surfaces and band dispersions covering the whole Brillouin zone by varying the incident photon
energy and the emission angle~\cite{Sekiyama2004a,Yokoya2005,Yano2007}.
Thanks to the high kinetic energy of the photoelectron, it is known that the probing depth of soft-x-ray
ARPES is longer than that of conventional VUV measurements~\cite{Tanuma1987}. Therefore it is a
suitable probe for the electronic states of the buried interfaces~\cite{Berner2013,Cancellieri2014} and the capped diluted magnetic semiconductors~\cite{Kobayashi2014}. Due to the recent
instrumental advance, it is also promising to reveal the three-dimensional band structure of the bulk
electronic states of strongly correlated oxides~\cite{Strocov2014}, which are often different from the
surface~\cite{Sekiyama2004b}. Especially, it has been reported that the valence band photoemission
spectra of strongly correlated vanadium oxide V$_2$O$_3$ show a very drastic development of the
prominent peak structure near the Fermi level ($E_{\textrm{F}}$) with increasing the photon energy~\cite{Mo2003,Fujiwara2011}, and thus high-energy photoemission is essential.

V$_2$O$_3$ shows a first order transition at $\sim$155 K from a high-temperature paramagnetic metal (PM) phase to a low-temperature antiferromagnetic insulator (AFI) phase, accompanied by a structural change from the corundum phase to the monoclinic phase. This metal-insulator transition (MIT) has been discussed as a paradigmatic example of the Mott-Hubbard MIT~\cite{Imada1998}, which is characterized by the ratio between the on-site Coulomb repulsive energy \textit{U} and one-electron bandwidth \textit{W}. However, the local density approximation (LDA) bandwidth \textit{W} with fixed \textit{U} cannot fully explain the MIT~\cite{Mo2003}, and the recent hard-x-ray photoemission reveals that \textit{U} does not change in the MIT. Polarized X-ray absorption spectroscopy has shown that the orbital population changes across the MIT~\cite{Park2000,Hansmann2012}. Thus, one needs new theoretical concepts such as the orbital selective MIT picture where the orbital degrees of freedom play an important role in the MIT~\cite{Laad2006,Poteryaev2007}. 

To reveal the driving mechanism of the MIT from the experimental side, it is essential to experimentally determine the band structure and Fermi surface topology of V$_2$O$_3$ for direct comparison with modern theories based on realistic models. Nevertheless, only a few works~\cite{Smith1988,Rodolakis2009} have been reported on the ARPES spectra because of the difficulty to obtain a high quality mirror plane after cleaving. The size of the flat regions on a cleaved surface has been reported to be typically no larger than the diameter of 100 $\mu$m~\cite{Mo2006}, and thus one must be able to select well cleaved regions with sub-mm$^2$ size to record reliable ARPES data in a wide momentum region by rotating the sample, which may induce possible change of the sample position. These technical problems are, however, overcome with micro positioning techniques for soft-X-ray ARPES measurements, namely, using focused soft-X-ray and monitoring the sample surface with a long working distance microscope for the microfocus beam (40 $\mu$m $\times$ 65 $\mu$m FWHMs)~\cite{Muro2009}. In this letter, we demonstrate successful band mapping and also show the position dependence of the photoemission spectral line shape for metallic V$_2$O$_3$. 

\section{Experimental}

\begin{figure}[t]
\begin{center}
\includegraphics[width=8cm,clip]{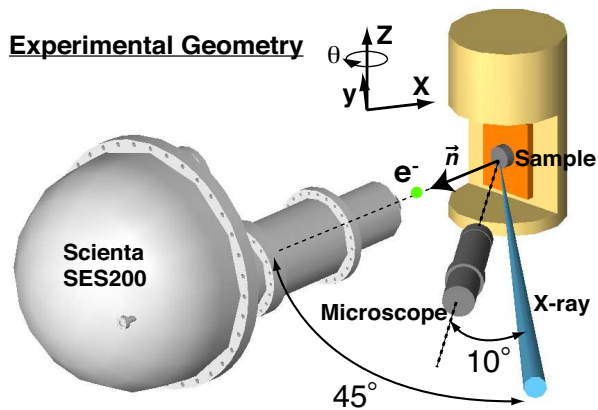}
\caption 
{Experimental geometry for the soft-X-ray ARPES demonstrated here. The sample is mounted on the xyz$\theta$-stage. $\vec{n}$ indicates the sample normal direction.}
\label{Fig1}
\end{center}
\end{figure}

\begin{figure}[t]
\begin{center}
\includegraphics[width=8cm]{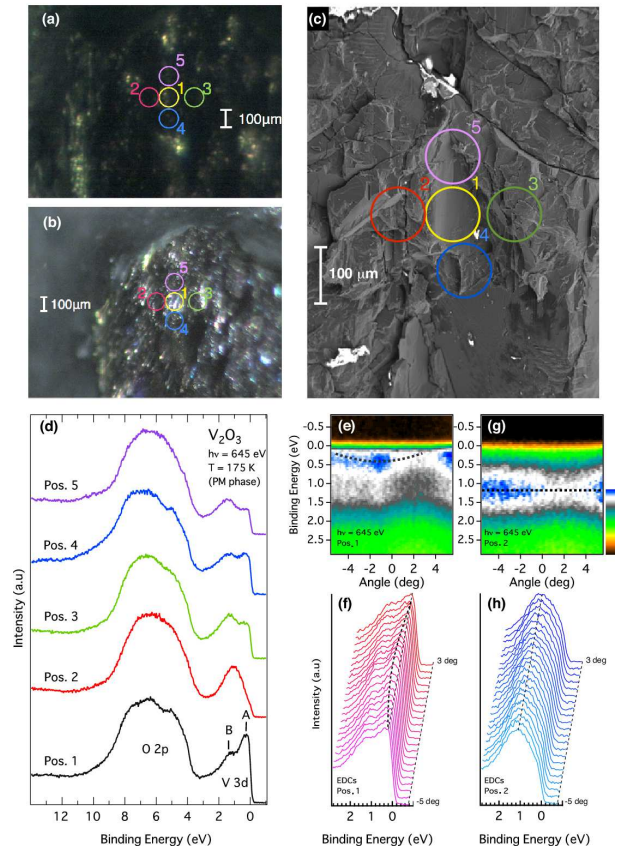}
\caption 
{The optical microscope images recorded in the main chamber (a) and the preparation chamber (b). (c) SEM image obtained in Osaka University after the photoemission measurements. Circles with diameter of 100 $\mu$m in (a-c) are indicators for the sampling points. (d) Sample position dependence of the angle-integrated valence band photoemission spectra recorded at the circles in (a), (b) and (d) with diameter of 100 $\mu$m. ARPES intensity plot (e) and the energy distribution curves (f) for position 1, and those for position 2 (g),(h). The dashed lines in (e-h) are guides to the eye of the peak position indicating the band structures.}
\label{Fig2}
\end{center}
\end{figure}

ARPES measurements were performed at the soft X-ray helical undulator beamline BL25SU~\cite{Saitoh1998,Saitoh2000} in SPring-8. The spectra were recorded with a hemispherical electron energy analyser (VG Scienta AB, SES200) with an angular resolution of 0.2${^\circ}$ within $\pm$5.5${^\circ}$ along the slit. In the experimental geometry for this mcro-ARPES measurements as described in Fig.~\ref{Fig1}, the entrance slit of the analyzer is vertical. Therefore, this electron energy analyzer has a horizontal spatial resolution when using the spatial imaging mode of the electron lens~\cite{Muro2009}. Since the magnification of the special imaging mode is 5, the horizontal spatial resolution is $\sim$40 $\mu$m for the 0.2-mm-width slit, and thus it is possible to select the small region on the sample surface. The analyzer is mounted at 45 degrees horizontally inclined from the soft-X-ray beam, which is well focused onto the sample surface with the beam size of 40 $\mu$m $\times$ 65 $\mu$m FWHMs. The long working distance microscope (Infinity Photo-Optical, K2/S) is mounted with an offset angle of 10 degree from the incident photon beam. The microscope images are displayed on a computer screen as shown in Fig.~\ref{Fig2}(a). The sample positions were selected by monitoring the microscope image of the sample surface and the photoelectron count rates. Beforehand, the intersection between the electron analyzer axis and the soft-X-ray beam was marked on the microscope monitor using a fluorescent substrate, which was positioned so as to maximize the photoelectron counts detected by the electron analyzer. To adjust a small area of the sample surface to this intersection, we first set the target region to the mark on the microscope monitor, and scanned the sample position along the microscope axis while keeping the target region on the mark until detecting the maximum photoelectron counts to avoid the ambiguity of the focal depth of the microscope. The detailed positioning techniques are described elsewhere~\cite{Muro2009,Muro2011}. The photoemission spectra are calibrated by the Fermi edge for evaporated Au. The total energy resolution at 175 K was set to 180$\sim$250 meV at $h{\nu}$ = 640 eV. 

\begin{figure*}
\begin{center}
\includegraphics[width=17cm]{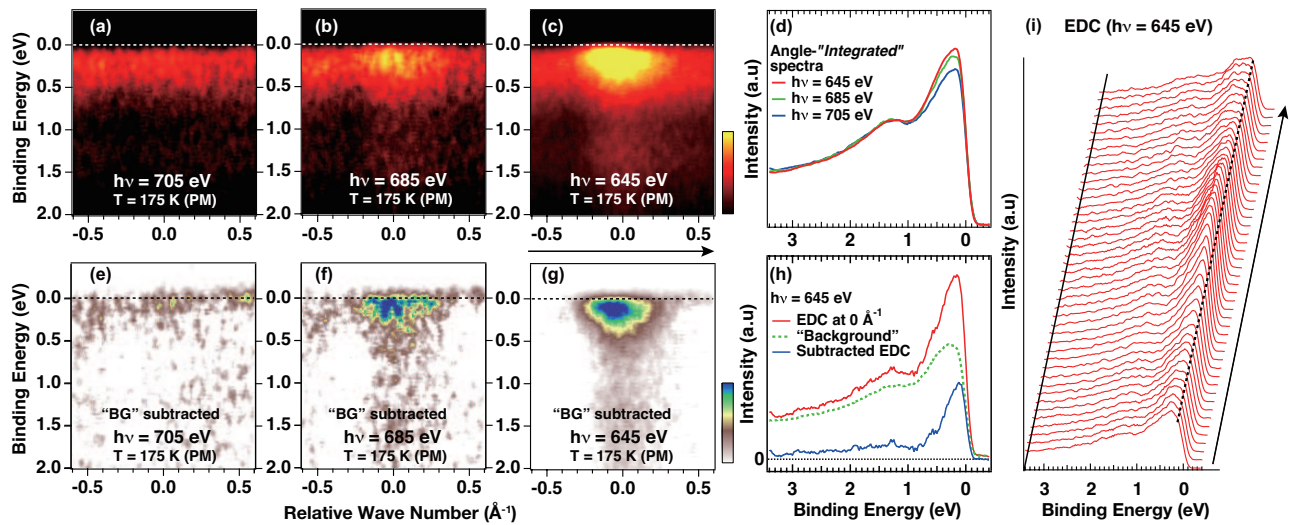}
\caption 
{Band mapping of V$_2$O$_3$ recorded at $h\nu$ = 705 eV (a)  685 eV(b), 645 eV (c), and those angle-\textit{`integrated'} spectra  (d). Highlighted band structures are shown by subtracting the non-dispersive background components in the range of -1 {\AA}$^{-1}$ to -0.5 {\AA}$^{-1}$ for $h\nu$ = 705 eV (e)  685 eV(f), 645 eV (g). This subtraction procedure is demonstrated in (h) for the energy distribution curve (EDC) at the relative wave number of 0 {\AA}$^{-1}$ recorded at $h\nu$ = 645 eV in (h).  (i) EDCs for (c) with the guides to the eye of dashed lines and the arrow.}
\label{Fig3}
\end{center}
\end{figure*}

The single crystalline V$_2$O$_3$ was cleaved in situ with a base pressure of $\sim$3$\times$10$^{-8}$ Pa. To cleave the crystal we glued a Cu plate to the V$_2$O$_3$ sample. With the sample temperature set somewhat below the MIT transition temperature, we then made thermal contact to the Cu plate with a room temperature diamond file that was mounted on a wobble stick, thus setting up a thermal gradient across the sample. After a waiting time that was variable from sample to sample, and with some adjustment of the starting temperature from sample to sample, this procedure causes the sample to cleave spontaneously due to the strain set up by the volume change that occurs across the MIT. We have another microscope to monitor the sample surface just after cleaving in the preparation chamber, wherefrom the sample is transferred to the main chamber using only the z(vertical)-motion~\cite{Muro2011}. The microscope in the preparation chamber gives better contrast of the image than that in the main chamber due to its shorter working distance, as shown under illumination in Fig.~\ref{Fig2}(b). Therefore, it is very helpful for selecting a well cleaved area on the cleaved sample surface. After the ARPES measurement, the surface roughness of the measured position was checked with an outside scanning electron microscope in Fig.~\ref{Fig2}(c), and/or with a confocal microscope in Osaka University. The effectively measured regions, taking into account the vibration of the manipulator ($\pm$15 $\mu$m) due to the closed-cycle cryostat~\cite{Muro2011}, are circled (diameter of 100 $\mu$m) on the microscope images in Figs.~\ref{Fig2}(a-c). 

\section{RESULTS AND DISCUSSION}

Figure~\ref{Fig2}(d) shows the angle integrated valence band photoemission spectra recorded at 5 different positions marked on the microscope image in Figs.~\ref{Fig2}(a) and 2(b). Although the distance between position 1 and other 4 positions is only $\sim$100 $\mu$m, the spectra show a clear sample position dependence. Most remarkable are the V $3d$ states in the range from $E_{\textrm{F}}$ to 3 eV. The spectrum observed at position 1 shows the peak structure labelled as A near $E_{\textrm{F}}$ and the shoulder structure B around 1.5 eV. The former peak is the so-called quasi particle peak, and the latter is due to the incoherent satellite originating from the lower Hubbard band. This spectral line shape is qualitatively consistent with the reported spectra~\cite{Mo2003}. The structures A and B are, however, not observed at position 2. One can find only a hump structure around 1 eV, and the O $2p$ states ( $4-10$ eV) are also rather featureless compared with those at position 1. The spectral line shapes obtained at the other 3 positions are somewhat in-between. These variations in the spectra could be due to the inhomogeneity of the electronic structure, as reported in recent absorption spectroscopy and photoelectron microscopy~\cite{Lupi2010}, and/or the increase of surface components due to the surface roughness.  Indeed, it is reported that the valence band spectrum excited by low energy photons ($\sim$60 eV) does not show the peak structure A~\cite{Mo2003}. By comparing the morphology on the SEM image in Fig.~\ref{Fig2}(c) to the valence band photoemission spectra, one can notice that only the flat surface in region 1 provides fine structured spectra. At other positions, the flat surface is coexistent with rough surfaces in the area within the diameter of 100 $\mu$m. Especially the position 2 even has a crack in the middle of the beam spot area. This suggests that the photoemission line shape of V$_2$O$_3$ is sensitive to the surface roughness. For the ARPES measurements, the surface roughness must be minimized because the flat surface is essential to enable the momentum conservation law. The ARPES spectra recorded in position 1 show dispersion near $E_{\textrm{F}}$ (Fig.~\ref{Fig2}(e)). The band width is not broad, but the electron like band feature is clearly observed in the energy distribution curves (EDCs) near $E_{\textrm{F}}$ in  Fig.~\ref{Fig2}(f). In contrast, the ARPES spectra obtained at position 2 show broad and rather dispersionless features in Figs.~\ref{Fig2}(g and h). These results show that selection of a proper cleavage plane with sub-mm$^2$ size by the micro positioning system is important for reliable ARPES measurements even when performing soft X-ray ARPES.

To discuss the electronic structures of three-dimensional materials such as V$_2$O$_3$, it is essential to obtain the band structure not only along $k_x$-$k_y$ but also in the $k_z$ directions. Since one can probe the band structure along the $k_z$ direction by sweeping the photon energy, we show in Fig.~\ref{Fig3} the $h\nu$-dependent ARPES spectra for the certain cleaved V$_2$O$_3$ surface as position 1 in Fig.~\ref{Fig2}(a). The band mapping recorded at different $h\nu$ in Figs.~\ref{Fig3}(a-c) clearly shows the development of the band structures with decreasing $h\nu$, representing the $k_z$ dispersion. It is obvious that there are no bands crossing $E_{\textrm{F}}$ in Fig.~\ref{Fig3}(a) recorded at $h\nu$ = 705 eV, while the V $3d$ bands contribute to the Fermi surface at $h\nu$ = 685 eV (Fig.~\ref{Fig3}(b)). The most prominent signals are observed at $h\nu$ = 645 eV (Fig.~\ref{Fig3}(c)). Indeed, the V $3d$ peak near $E_{\textrm{F}}$ in the angle-\textit{'integrated'} spectra increases with decreasing $h\nu$ from 705 eV to 645 eV (Fig.~\ref{Fig3}(d)). This evolution of the band structure is highlighted in Figs.~\ref{Fig3}(e-g) by subtracting the partially momentum averaged EDC in the range of -1 {\AA}$^{-1}$ to -0.5 {\AA}$^{-1}$ as shown in Fig.~\ref{Fig3}(h), where the non-dispersive signal is observed, to give a `{background}Õ signal possibly due to the angle-\textit{`integrated'} components from the surface roughness~\cite{Mo2006} and/or the thermal effects because of high excitation energies combined with relatively high sample temperature~\cite{Kamakura2006,Braun2013}. Moreover, one can estimate the V $3d$ band width as $\sim$0.5 eV from the bottom of the band observed using 645 eV-photons. This is also supported by the EDCs in Fig.~\ref{Fig3}(i). By tracking the EDC peak positions near $E_{\textrm{F}}$, one notices the electron-like band dispersion. 

\begin{figure}
\begin{center}
\includegraphics[width=8cm]{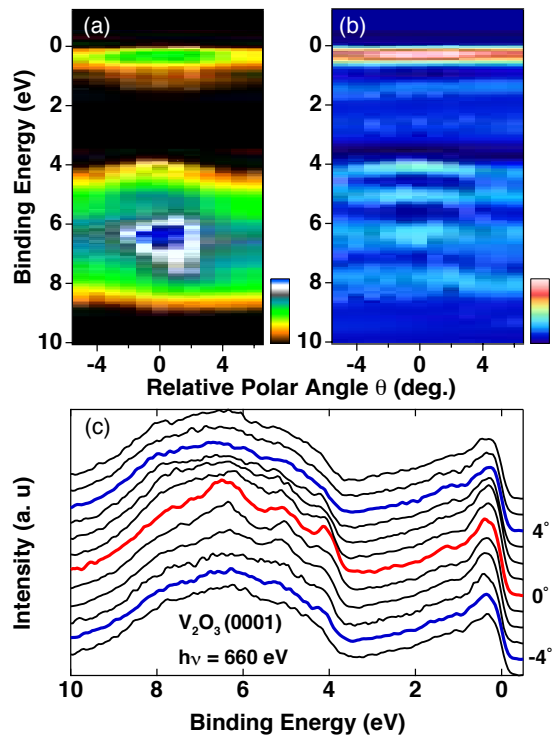}
\caption 
{Polar angle dependence of ARPES recorded with 660 eV-photons (a) and second derivative (b) for oriented V$_2$O$_3$ with (0001) plane. EDCs of (a) is shown in (c). }
\label{Fig4}
\end{center}
\end{figure}

To demonstrate the feasibility of the three-dimensional ARPES, we have measured the polar angle dependence of the ARPES spectra by rotating the manipulator $\theta$ for \textit{`oriented'} V$_2$O$_3$ cleaved on the (0001) plane.  The intensity plot of the polar angle dependent ARPES spectra with $h\nu$ = 660 eV  in Fig.~\ref{Fig4}(a) shows a narrow band structure around $E_{\textrm{F}}$, and the top of the O $2p$ band is observed at the relative angle of 0 degree. By taking the second derivative as shown in Fig.~\ref{Fig4}(b), the O $2p$ band in the range of $4-8$ eV is clearly observed. Especially three O $2p$ bands are confirmed by tracking the peak of the EDCs around 0 degree in Fig. 4(c). Furthermore one recognizes that the bottom of the V $3d$ band is located at 0 degree and weekly disperses to $\pm$4 degrees. Thus the feasibility of the polar angle dependent ARPES for V$_2$O$_3$ is shown, and we now stress that three-dimensional ARPES for poorly cleaving strong correlated systems, e.g., lacking a layer structure, is feasible by selecting the tiny flat surfaces using the micro-positioning technique.

\section{CONCLUSION}

We have succeeded in recording band dispersions of metallic V$_2$O$_3$ by using a micro soft-X-ray ARPES method that combines a focused X-ray beam and the micro-positioning techniques with a long distance optical microscope. The spectra showed a strong position dependence with respect to surface roughness, and band mapping was successfully performed in a flat region of size $\sim$100 $\mu$m. The photon energy dependence of ARPES measurements has shown the evolution of the V $3d$ band structures, and the polar angle dependence has captured the symmetric band dispersions in the O $2p$ states. Therefore, we conclude that complete three-dimensional band mapping and Fermiology along all $k_x$, $k_y$, $k_z$ directions are feasible by performing soft-X-ray ARPES with micro positioning techniques to select a micro cleavage surface. 

\section{ACKNOWLEDGMENTS}

The ARPES measurements were supported by K. Yamagami, S. Naimen and T. Matsushita. This work was supported by MEXT/JSPS KAKENHI Grant Number 23740240, and the Grant-in-Aid for Innovative Areas (20102003) ``Heavy Electrons'' from MEXT Japan, as well as by Toray Science Foundation. The measurements were under the approval of the Japan Synchrotron Radiation research Institute (2011B1348, 2012A1486 and 2013A1089).


\end{document}